\documentclass[preprintnumbers,article,amsmath,amssymb,floatfix,10pt,prd,onecolumn,
superscriptaddress,nofootinbib]{revtex4-2}
\usepackage{bm}
\usepackage{amsfonts}
\usepackage{latexsym}
\usepackage[latin1]{inputenc}
\usepackage{graphicx}
\usepackage{amsmath}
\usepackage{palatino}
\usepackage{mathpazo}
\usepackage{textcomp}
\usepackage{float}
\usepackage{booktabs}
\usepackage{dcolumn}
\usepackage{ragged2e}
\usepackage{hyperref}
\hypersetup{colorlinks,citecolor=blue}
\hypersetup{colorlinks=true,linkcolor=blue,filecolor=magenta,    urlcolor=blue}
\usepackage{amsmath}
\usepackage{subfigure}
\usepackage[]{natbib}
\usepackage{xcolor}
\usepackage{orcidlink}
\usepackage{epsfig}
\usepackage{caption}
\usepackage{subcaption}
\usepackage{commath}
\def\jnl@style{\it}
\def\aaref@jnl#1{{\jnl@style#1}}

\def\aaref@jnl#1{{\jnl@style#1}}

\def\aj{\aaref@jnl{AJ}}                   % Astronomical Journal
\def\apj{\aaref@jnl{ApJ}}                 % Astrophysical Journal
\def\apjl{\aaref@jnl{ApJ}}                % Astrophysical Journal, Letters
\def\apjs{\aaref@jnl{ApJS}}               % Astrophysical Journal, Supplement
\def\apss{\aaref@jnl{Ap\&SS}}             % Astrophysics and Space Science
\def\aap{\aaref@jnl{A\&A}}                % Astronomy and Astrophysics
\def\aapr{\aaref@jnl{A\&A~Rev.}}          % Astronomy and Astrophysics Reviews
\def\aaps{\aaref@jnl{A\&AS}}              % Astronomy and Astrophysics, Supplement
\def\mnras{\aaref@jnl{Mon.~Not.~Roy.~Astron.~Soc.}}             % Monthly Notices of the RAS
\def\prd{\aaref@jnl{Phys.~Rev.~D}}        % Physical Review D
\def\prc{\aaref@jnl{Phys.~Rev.~C}}  % Physical Review C
\def\prl{\aaref@jnl{Phys.~Rev.~Lett.}}    % Physical Review Letters
\def\qjras{\aaref@jnl{QJRAS}}             % Quarterly Journal of the RAS
\def\skytel{\aaref@jnl{S\&T}}             % Sky and Telescope
\def\ssr{\aaref@jnl{Space~Sci.~Rev.}}     % Space Science Reviews
\def\zap{\aaref@jnl{ZAp}}                 % Zeitschrift fuer Astrophysik
\def\nat{\aaref@jnl{Nature}}              % Nature
\def\aplett{\aaref@jnl{Astrophys.~Lett.}} % Astrophysics Letters
\def\apspr{\aaref@jnl{Astrophys.~Space~Phys.~Res.}} % Astrophysics Space Physics Research
\def\physrep{\aaref@jnl{Phys.~Rep.}}      % Physics Reports
\def\physscr{\aaref@jnl{Phys.~Scr}}       % Physica Scripta
\def\commat{\aaref@jnl{Comm.~Math.~Phys.}}              % Communications in Mathematical Physics
\def\science{\aaref@jnl{Science}}               % Science
\def\cqg{\aaref@jnl{Classical Quant.~Grav.}}            % Classical and Quantum Gravity
\def\jpcs{\aaref@jnl{JPCS}}                                     % Journal of Physics Conference Series
\def\ijmpd{\aaref@jnl{Int.~J.~Mod.~Phys.~D}}                    % International Journal of Modern Physics D
\def\grg{\aaref@jnl{Gen.~Relat.~Gravit.}}               % General Relativity and Gravitation
\def\rpp{\aaref@jnl{Rep.~Prog.~Phys.}}          % Reports on Progress in Physics
\def\npa{\aaref@jnl{Nucl.~Phys.~A}}        % Nuclear Physics A
\def\lrr{\aaref@jnl{Living Rev.~Rel.}}                   % Living reviews in relativity
\def\jcap{\aaref@jnl{J.~Cosmology Astropart.~Phys.}}    % Journal of cosmology and astroparticle physics
\def\rmp{\aaref@jnl{Rev.~Mod.~Phys.}}   %Reviews of modern physics
\def\epjc{\aaref@jnl{Eur.~Phys.~J.~C}} 
\def\plb{\aaref@jnl{~Phy.~Lett.~B}} 
\def\mpla{\aaref@jnl{Mod.~Phy.~Lett.~A}} 
\def\arxiv{\aaref@jnl{arxiv.org}}

\allowdisplaybreaks[1]
\begin{document}
%\color{red}
\color{black}       %% For one column
\title{Signature flips in time-varying $\Lambda(t)$ cosmological models with observational data}

\author{Yerlan Myrzakulov\orcidlink{0000-0003-0160-0422}}\email[Email: ]{ymyrzakulov@gmail.com \textcolor{black}{(corresponding author)}} 
\affiliation{Department of General \& Theoretical Physics, L.N. Gumilyov Eurasian National University, Astana, 010008, Kazakhstan.}
\affiliation{Ratbay Myrzakulov Eurasian International Center for Theoretical Physics, Astana, 010009, Kazakhstan.}

\author{M. Koussour\orcidlink{0000-0002-4188-0572}}
\email[Email: ]{pr.mouhssine@gmail.com}
\affiliation{Department of Physics, University of Hassan II Casablanca, Morocco.}

\author{M. Karimov}
\email[Email: ]{karimovmuzaffar050@gmail.com}
\affiliation{Faculty of Mathematics, Namangan State University, Boburshoh str. 161, Namangan 160107, Uzbekistan.}

\author{J. Rayimbaev\orcidlink{0000-0001-9293-1838}}
\email[Email: ]{javlon@astrin.uz}
\affiliation{New Uzbekistan University, Mustaqillik Ave. 54, Tashkent 100007, Uzbekistan.}
\affiliation{University of Tashkent for Applied Sciences, Gavhar Str. 1, Tashkent 100149, Uzbekistan.}
\affiliation{National University of Uzbekistan, Tashkent 100174, Uzbekistan.}

%%%%%%%%%%%%%%%%%%%%%%%%%%%%%%%%%%%%  DATE  %%%%%%%%%%%%%%%%%%%%%%%%%%%%%%%%%%%%
\date{\today}
\begin{abstract}
In this study, we investigate signature flips within the framework of cosmological models featuring a time-varying vacuum energy term $\Lambda(t)$. Specifically, we consider the power-law form of $\Lambda=\alpha H^n$, where $\alpha$ and $n$ are constants. To constrain the model parameters, we use the MCMC technique, allowing for effective exploration of the model's parameters. We apply this approach to analyze 31 points of observational Hubble Data (OHD), 1048 points from the Pantheon data, and additional CMB data. We consider three scenarios: when $n$ is a free parameter (Case I), when $n=0$ (Case II), and when $n=1$ (Case III). In our analysis across all three cases, we observe that our model portrays the universe's evolution from a matter-dominated decelerated epoch to an accelerated epoch, as indicated by the corresponding deceleration parameter. In addition, we investigate the physical behavior of total energy density, total EoS parameter, and jerk parameter. Our findings consistently indicate that all cosmological parameters predict an accelerated expansion phase of the universe for all three cases ($q_0<0$, $\omega_0<-\frac{1}{3}$, $j_0>0$). Furthermore, our analysis reveals that the $Om(z)$ diagnostics for Cases I and III align with the quintessence region, while Case II corresponds to the $\Lambda$CDM model. 


\end{abstract}

\maketitle

\tableofcontents

\section{Introduction}\label{sec1}

Recent astronomical observations have provided strong evidence supporting the theory of accelerated expansion in our universe \cite{Riess/1998,Perlmutter/1999,Komatsu/2011,Planck/2014,Eisenstein2005,Percival/2007,Farooq/2017,Yu_2018}. This acceleration is attributed to an unknown form of matter known as dark energy (DE), characterized by its negative pressure and positive energy density, satisfying the condition $\rho+3p<0$ \cite{B1,C1,Brevik}. The nature of DE remains one of the most significant mysteries in modern cosmology, driving ongoing research and theoretical investigations into its properties and implications for the evolution of the universe \cite{Steinhardt/2011}. Recent studies suggest that the composition of the universe is approximately 74\% DE, 22\% dark matter, and 4\% ordinary matter. Among the various forms of DE, the simplest is known as the cosmological constant ($\Lambda$), which is a key component of the $\Lambda$ cold-dark-matter ($\Lambda$CDM) model \cite{TP1,VS1}. The $\Lambda$CDM model has proven to be successful in describing various observed features of the universe, providing a framework that aligns with a wide range of astronomical observations. This model has become a cornerstone of modern cosmology and serves as the foundation for many theoretical and observational studies.

While the $\Lambda$CDM cosmological model aligns well with a majority of observational data, it encounters significant challenges, particularly regarding fine-tuning \cite{Weinberg/1989}. Observations indicate that the value of $\Lambda$ is exceptionally small, just enough to account for the current acceleration of the universe. However, theoretical predictions based on quantum field theory suggest a vastly larger value for $\Lambda$, differing from the observed value by a staggering factor of $10^{120}$ . Another issue, known as the coincidence problem \cite{Steinhardt/1999}, arises from the unexpectedly similar magnitudes of $\Lambda$ and ordinary matter in the universe. This coincidence raises questions about the underlying reasons for the comparable scales of these two components at the present epoch. These challenges have motivated researchers to explore alternative explanations and modifications to the $\Lambda$CDM model to address these issues and provide a more comprehensive understanding of the universe's dynamics. In the exploration of the universe's accelerated expansion, researchers have investigated various cosmological models using two primary approaches. One approach involves modifications to classical general relativity (GR), leading to modified theories of gravity \cite{H.A.,H.K.,Odintsov1,Odintsov2,T1,T2,T3,T4,Q0,Q1,Q2,Q3,Q4,Q5,Q6,Q7,Q8,Q9,Q10}. The other approach involves the introduction of mysterious energy components, commonly referred to as DE. In this context, the equation of state (EoS) parameter, which represents the ratio of pressure to energy density, plays a crucial role. Observationally, the EoS parameter is found to be very close to -1, indicating that the universe is in an accelerating phase, with the EoS parameter satisfying $\omega <-\frac{1}{3}$. To address the nature of dark energy, numerous studies have explored various theoretical frameworks, including quintessence \cite{RP}, phantom \cite{M.S.,M.S.-2}, K-essence \cite{T.C.,C.A.}, Chaplygin gas \cite{M.C.,A.Y.}, tachyon \cite{T.P.}, and holographic DE model \cite{F2,S10,M2}.

On the other hand, different models featuring a time-varying $\Lambda(t)$ or vacuum decay have been proposed in different contexts, often assuming time dependencies for $\Lambda(t)$ \cite{Oztas,Vishwakarma,Overduin}, where different phenomenological decay laws for $\Lambda(t)$ have been explored. The expression for $\Lambda(t)$ can also be obtained through geometric analysis \cite{Azri1,Azri2} or quantum mechanical reasoning \cite{Szydlowski}. Interactions between vacuum and matter have been studied in various approaches, with comparisons made to recent cosmological data \cite{Bruni,Papagiannopoulos,Benetti1,Benetti2}. One promising approach to address the shortcomings of the $\Lambda(t)$CDM model is the running vacuum model (RVM). This model arises from applying the renormalization group approach of quantum field theory in curved spaces to renormalize the vacuum energy density. The evolution of the vacuum energy density in this model is expressed as a series of powers of the Hubble function $H$ and its derivatives for cosmic time. While the leading term of this expansion is constant, the next-to-leading term evolves as $H^2$, which can have implications for the current evolution of the scale factor. Initially introduced in a semi-qualitative manner, the RVM has more recently been rigorously derived from quantum field theory in curved spacetime. This class of models has shown promise in fitting with cosmological observables, potentially outperforming the $\Lambda(t)$CDM model in certain scenarios \cite{Sola2017, Peracaula2018, Tsiapi2019, Mavromatos2021, Peracaula2023, Peracaula2021, Peracaula2018b}. For more details on the RVM and recent theoretical developments, interested readers can refer to the relevant literature \cite{Pulido2020,Pulido2022a,Pulido2022b,Pulido2023,Peracaula2022}. Overduin and Cooperstock \cite{Overduin} explored the evolution of the scale factor in the context of a variable vacuum energy term, such as $\Lambda = A t^{-l}$, $\Lambda = B a^{-m}$, $\Lambda = C t^{n}$ and $\Lambda = D q^{r}$ (where $A$, $B$, $C$, $D$, $l$, $m$, $n$, $r$ are constants). In a recent study by Rezaei et al. \cite{Rezaei}, the authors employed phenomenological reasoning to parameterize the time-dependent behavior of $\Lambda(t)$, expressing it as a power series expansion of the Hubble rate and its time derivatives: $\Lambda(t) \propto H, \Lambda(t) \propto \dot{H}, \Lambda(t) \propto H^2$. In addition, Singh and Sol\`{a} \cite{Singh/2021} investigated Friedmann cosmology within the framework of the Brans-Dicke theory, considering a scenario where the vacuum density decays over time. Specifically, the authors explored $\Lambda(t)$ given by the phenomenological law $\Lambda(t)=\lambda+\sigma H$, where $\lambda$ and $\sigma$ are constants and $H$ is the Hubble parameter. Khatri and Singh \cite{Khatri/2023, Singh/2023}, within the framework of Brans-Dicke theory, focused on constraining models of time-varying vacuum energy. They introduced two forms for the cosmological constant: a power-series form and a power-law form.

In this work, we explore signature flips in cosmological models with a time-varying vacuum energy term $\Lambda(t)$. These flips represent changes in the behavior of cosmological parameters, especially the transition from decelerated to accelerated expansion. We focus on the power-law form of $\Lambda=\alpha H^n$, where $\alpha$ and $n$ are constants \cite{Overduin}. Specifically, we consider three scenarios: when $n$ is a free parameter (Case I), when $n=0$ (Case II), and when $n=1$ (Case III). To refine our model's parameters, we use recently updated datasets, which include 31 data points from the OHD dataset and 1048 data points from the Pantheon dataset \cite{Yu_2018,Scolnic_2018}. In addition, we incorporate CMB data from Planck's recent results \cite{Planck/2020}. The paper is structured as follows. In Sec. \ref{sec2}, we introduce the flat FLRW Universe within the framework of $\Lambda(t)$CDM cosmology. In Sec. \ref{sec3}, we derive the expressions for the Hubble parameter under the assumption of the power-law form of $\Lambda(t)$. Subsequently, in Sec. \ref{sec4}, we present the best-fit values of the model parameters obtained using the combined OHD+Pantheon+CMB dataset for all three cases of $n$. In addition, we explore the behavior of the deceleration parameter, total energy density, total EoS parameter, jerk parameter, and the $Om(z)$ diagnostics in Sec. \ref{sec5}. Finally, in Sec. \ref{sec6}, we summarize the outcomes of our investigation.  

\section{$\Lambda(t)$CDM cosmology}\label{sec2}

The cosmological principle posits that, at a large scale, the universe is uniform in its distribution of matter (homogeneous) and looks the same in all directions (isotropic). This principle forms the cornerstone of modern cosmology, serving as a fundamental framework for comprehending the structure and development of the universe. Over the past few decades, the cosmological principle has been subjected to extensive testing using diverse methods, including observations of cosmic microwave background radiation, large-scale galaxy surveys, and studies of the distribution of matter in the universe \cite{Komatsu/2011,Planck/2014}. The cosmological principle is mathematically formulated by the flat Friedmann-Lema\^itre-Robertson-Walker (FLRW) metric,
\begin{equation} \label{FLRW}
ds^2= -dt^2 + a^2(t)[dx^2+dy^2+dz^2], 
\end{equation}
where $a(t)$ represents the scale factor of the universe.

In the context of the metric (\ref{FLRW}), the total energy-momentum tensor for a perfect fluid distribution, encompassing all contributions from various energy components in the universe, including matter and vacuum energy, is expressed as
\begin{equation}
\mathcal{T}_{\mu \nu }=(\rho_{m}+\rho_{\Lambda} +p_{m}+p_{\Lambda})u_{\mu }u_{\nu }+(p_{m}+p_{\Lambda})g_{\mu \nu },  \label{EMT}
\end{equation}%
where $\rho_m$ is the energy density of matter in the universe, $\rho_{\Lambda}$ is the vacuum energy density term, $p_{m}$ is the pressure of matter, $p_{\Lambda}$ is the pressure of the vacuum energy term, $u^{\mu}=(1,0,0,0)$ is the 4-velocity of the perfect fluid, describing its motion in spacetime, and $g_{\mu \nu}$ is the metric tensor of the FLRW spacetime, describing the geometry of the universe.

By using the Einstein field equations with respect to Eq. (\ref{FLRW}) in the presence of matter modeled as a perfect fluid (\ref{EMT}), we derive the Friedmann equations, which govern the dynamics of the universe and are expressed in terms of the Hubble parameter $(H=\frac{\dot a}{a})$ as \cite{Myrzakulov/2023}
\begin{eqnarray}
\label{F1}
3H^{2}&=&\rho_{m}+\rho_{\Lambda}, \\
2{\dot{H}}+3H^{2}&=&-(p_{m}+p_{\Lambda}). 
   \label{F2} 
\end{eqnarray}

The choice of units has been made so that the constants $8 \pi G=c=1$. In the present work, our focus is on the late-time universe, where we disregard the influence of radiation, i.e. $\rho_{r}=0$. On the other hand, $\rho_{\Lambda}$ and $p_{\Lambda}$ correspond to the density and pressure associated with the vacuum energy term, which can vary with time $\Lambda(t)$. In addition, we adopt the equation of state (EoS) of vacuum energy as $\omega_{\Lambda}=-1$ \cite{Macedo/2023}, which implies that the pressure of the vacuum energy term $p_{\Lambda}=-\Lambda$ is equal to minus its energy density $\rho_{\Lambda}=\Lambda$, as per the equation $p_{\Lambda}=-\rho_{\Lambda}$. By applying the divergence operator to Eq. (\ref{EMT}), we can derive the continuity equation for the matter component of the universe as follows:
\begin{equation}
 \dot{\rho}_{m} + 3H (\rho_{m}+p_{m}) = -\dot{\Lambda}, 
\end{equation}
which indicates that matter is not conserved independently, as the decaying vacuum serves as a source of matter. By using the first Friedmann equation (\ref{F1}), we can derive the expression for the energy density of matter as $\rho_{m}=3 H^2 -\Lambda$. Now, the present value of the dimensionless density parameters corresponding to matter and the vacuum energy term are defined as
\begin{equation}\label{3j}
  \Omega_{m0}=\frac{\rho_{m0}}{3H_{0}^2}, \ \ \ \Omega_{\Lambda 0}=\frac{\Lambda_0}{3H_{0}^2}, \ \  \   \Omega_{m0}+\Omega_{\Lambda 0}=1,
\end{equation}
where $\rho_{m0}$ is a constant of integration that represents the current value of the matter density.

Then, we can derive the expression for the total EoS parameter as 
\begin{equation}
    \omega_{tot}=\frac{p_{\Lambda}}{\rho_{m}+\rho_{\Lambda}}=-\frac{1}{1+\frac{\rho_m}{\rho_{\Lambda}}},
    \label{EoS_tot}
\end{equation}
where $\omega_{\Lambda}=-1$. Using the assumption of pressureless matter ($p_m=0$), the second Friedmann equation (\ref{F2}) can be expressed as \cite{Macedo/2023,Koussour/2024}
\begin{equation} \label{F22}
2{\dot{H}}+3H^{2}=\Lambda.
\end{equation}

Furthermore, to establish cosmological constraints, we choose the redshift $z$ as the independent variable instead of the conventional time variable $t$. The redshift $z$ is defined by the formula: $1+z=\frac{1}{a(t)}$, where $a_0$ is the scale factor at present $(a(0) = 1)$ and $a(t)$ is the scale factor at the time corresponding to the redshift $z$. Using $z$ as the independent variable allows us to directly relate observational data, such as the redshift of distant galaxies, to the evolution of the universe. Hence, we can replace time derivatives with derivatives with respect to redshift using the following relation:
\begin{equation}
\label{dt}
    \frac{d}{dt}=-H(z)(1+z) \frac{d}{dz}.
\end{equation}

Now, using the definition of the derivative of the Hubble parameter with respect to time, Eq. (\ref{F22}) becomes
\begin{equation} \label{EHP}
   \frac{dH(z)}{dz}=\frac{3H(z)}{2(1+z)}-\frac{\Lambda(z)}{2H(z)(1+z)}. 
\end{equation}

\section{Cosmological $\Lambda(t)$CDM models} \label{sec3}

In this section, we will analyze the results obtained in the previous section for several $\Lambda(t)$CDM models. Here, we consider the following $\Lambda(t)$CDM model, which incorporates a power-law form of the Hubble parameter, as described in \cite{Overduin},
\begin{equation} \label{Lz}
    \Lambda= \alpha H^n,
\end{equation}
where $\alpha$ and $n$ are the arbitrary constants. For this particular selection of the function $\Lambda$, we can derive the following differential equation:
\begin{equation} \label{EHP}
   \frac{dH(z)}{dz}=\frac{3H(z)}{2(1+z)}-\frac{\alpha H(z)^{(n-1)}}{2(1+z)}. 
\end{equation}

Now, by using Eqs. (\ref{3j}) and (\ref{Lz}) we have
\begin{equation}
  \Omega_{\Lambda 0}=\frac{\alpha}{3H_{0}^{(2-n)}}.
\end{equation}

After integrating the aforementioned equation, we derive the following expression for the Hubble parameter $H(z)$ in terms of redshift, which applies specifically to the $\Lambda(t)$CDM model:
\begin{eqnarray}
H\left( z\right) ^{\left( 2-n\right) } &=&H_{0}^{\left( 2-n\right) }\left(
1+z\right) ^{\gamma }-\frac{\alpha }{3}\left( 1+z\right) ^{\gamma }+\frac{%
\alpha }{3},  \nonumber \\
&=&H_{0}^{\left( 2-n\right) }\left( 1+z\right) ^{\gamma }-H_{0}^{\left(
2-n\right) }\Omega _{\Lambda 0}\left( 1+z\right) ^{\gamma }+H_{0}^{\left(
2-n\right) }\Omega _{\Lambda 0},  \nonumber \\
&=&H_{0}^{\left( 2-n\right) }\left[ \left( 1+z\right) ^{\gamma }-\Omega
_{\Lambda 0}\left( 1+z\right) ^{\gamma }+\Omega _{\Lambda 0}\right],  
\nonumber \\
&=&H_{0}^{\left( 2-n\right) }\left[ \left( 1-\Omega _{\Lambda 0}\right)
\left( 1+z\right) ^{\gamma }+\Omega _{\Lambda 0}\right],   \nonumber \\
&=&H_{0}^{\left( 2-n\right) }\left[ \Omega _{m0}\left( 1+z\right) ^{\gamma
}+\left( 1-\Omega _{m0}\right) \right],   \nonumber \\
H\left( z\right)  &=&H_{0}\left[ \Omega _{m0}\left( 1+z\right) ^{\gamma
}+\left( 1-\Omega _{m0}\right) \right] ^{\frac{1}{\left( 2-n\right) }},
\label{Hz}
\end{eqnarray}
where $H_0$ is the present value of the Hubble parameter and $\gamma=3-\frac{3 n}{2}$. 

Now, we consider three different cases related to the vacuum energy term $\Lambda(t)$, which are well-documented in the existing literature. Each case represents a distinct scenario that can significantly impact our understanding of the universe's evolution within the context of the $\Lambda(t)$CDM model.

\begin{itemize}
    \item Case I: When $n$ is a free parameter in the $\Lambda(t)$CDM model. In this scenario, we need to constrain three model parameters: $H_0$, $\Omega_{m0}$, and $n$. These parameters are crucial for understanding the dynamics of the universe and can be constrained through observational data and theoretical considerations.
    \item Case II: When $n=0$, the scenario corresponds to $\Lambda=\alpha>0$, which aligns with the standard $\Lambda$CDM model. In this case, the expression for the Hubble parameter is given by
    \begin{equation}
      H\left( z\right)=H_{0}\left[ \Omega _{m0}\left( 1+z\right) ^{3
}+\left( 1-\Omega _{m0}\right) \right] ^{\frac{1}{2}}  
    \end{equation}
    Specifically, when $\alpha=0$,  the expression for $H(z)$ reduces to $H(z)=H_{0}(1+z)^{\frac{3}{2}}$, which corresponds to a matter-dominated universe.
    \item Case III: When $n=1$, this scenario aligns with the linear model discussed in \cite{Rezaei}. In this case, the expression for the Hubble parameter is given by
    \begin{equation}
        H(z)=H_{0} \left[\sqrt{\Omega_{m0}^2 (1+z)^3}+(1-\Omega_{m0})\right].
    \end{equation}

\end{itemize}

\section{Observational constraints and methodology} \label{sec4}

In this analysis, we use three distinct datasets: observational measurements of the Hubble parameter (OHD), data on the distance modulus of Type Ia supernovae (SNe), and cosmic microwave background (CMB) data. These datasets have been employed to limit the model parameters $\Theta=(H_0,\Omega_{m0},n)$. For this reason, we use statistical analysis methods to perform numerical analysis, specifically employing the Markov Chain Monte Carlo (MCMC) method to obtain posterior distributions of the parameters. This analysis is implemented using the Python package \texttt{emcee} \citep{emcee}.

The best-fit parameter values are found by maximizing the likelihood function $\mathcal{L} = exp(-\chi^2/2)$, where $\chi^2$ represents the chi-squared function. The specific forms of the $\chi^2$ functions for different datasets are below. In addition, in all our analyses, we employ the following prior distributions for the parameters:
\begin{equation}
\begin{gathered}
H_0 \in [60,80] \text{km/s/Mpc}, \quad \Omega_{m0} \in [0,1], \quad n \in [-10,10].\\
\end{gathered}
\end{equation}

\subsection{OHD}
The observational Hubble data (OHD) is widely recognized as an important tool for studying the expansion history of the universe, especially beyond the framework of GR and the $\Lambda$CDM model. The OHD dataset is primarily derived using the differential age of galaxies method \cite{Yu_2018}. In our study, we focus on a compilation consisting of 31 data points from the Cosmic Chronometers (CC). To constrain our model, we introduce the chi-square function, which is commonly used to quantify the goodness of fit between theoretical predictions and observational data,
\begin{equation}
\chi^2_{OHD}=\sum_{i=1}^{31}\bigg[\frac{H^\mathrm{th}_i(\Theta,z_i)-H^\mathrm{obs}_i(z_i)}{\sigma_{H(z_i)}}\bigg]^2,
\end{equation}
where $H_{obs}$ and $H_{th}$ denote the observed and theoretical values of the Hubble parameter, respectively, while $\sigma_{H(z_{i})}$ denotes the standard deviation of the $i_{th}$ data point.

\subsection{Pantheon}
At first, observational studies on a select "golden sample" of 50 Type Ia SNe indicated that our universe is undergoing an accelerated phase of expansion. Following these initial findings, studies on SN datasets have expanded significantly over the past two decades, encompassing increasingly larger samples. Recently, a new sample of Type Ia SNe datasets has been released, comprising 1048 data points. In this article, we use the Pantheon datasets \cite{Scolnic_2018,Chang_2019}, which consist of 1048 spectroscopically confirmed Type Ia SNe samples covering the redshift range $z \in [0.01,2.3]$. 

The chi-square function for the Pantheon datasets is defined as 
\begin{equation}
    \chi^2_{Pantheon}= \sum^{1048}_{i,j=1}\frac{\Delta \mu_i}{\sigma_{\mu(z_i)}}, \quad  \Delta\mu_i=\mu^{th}(\Theta,z_i)-\mu_i^{obs}(z_i).
\end{equation}

The theoretical distance modulus, which is the logarithmic representation of distance, is expressed as
\begin{equation}
\mu^{th}=5\log_{10}d_L(z)+\mu_0,\quad \mu_0 = 5 \log_{10}(1/H_{0}Mpc)+25.
\end{equation}

The luminosity distance is defined as the measure of the amount of light received from a distant source, taking into account the inverse square law for light propagation in an expanding universe. For a spatially flat universe, it is defined as
\begin{equation}
d_L(z)=c(1+z)\int^z_0\frac{d\overline{z}}{H(\overline{z})},
\end{equation}
where $c$ is the speed of light.

In addition, the nuisance parameters in the Tripp formula \citep{tripp1998two}, denoted by $\mu= m_{B}-M_{B}+\alpha x_{1}-\beta c+ \Delta_{M}+\Delta_{B}$ were determined using the BEAMS with Bias Correction (BBC) method, as introduced by \citep{kessler2017}. Thus, the observed distance modulus is calculated as the discrepancy between the adjusted apparent magnitude $M_{B}$ and the absolute magnitude $m_{B}$ i.e. $\mu = m_{B} - M_{B}$. Therefore, the chi-square function can be represented as \citep{Deng2018}
\begin{equation}
\chi^2_{\mathrm{Pantheon}}=\Delta \mu^T C^{-1} \Delta \mu,
\end{equation}
where $C$ represents the total uncertainty matrix of the distance modulus, defined as
\begin{equation}
C = D_{\mathrm{stat}}+C_{\mathrm{sys}}
\end{equation}

In this context, we assume that the diagonal matrix of statistical uncertainties takes the form $D_{\mathrm{stat},ii}=\sigma^2_{\mu(z_i)}$. Additionally, we obtain systematic uncertainties using the BBC method described in Scolnic et al. \cite{Scolnic_2018}.
\begin{equation}
C_{ij,\mathrm{sys}} = \sum^K_{k=1}\bigg(\frac{\partial \mu^{obs}_i}{\partial S_k}\bigg)
\bigg(\frac{\partial \mu^{obs}_j}{\partial S_k}\bigg)\sigma^2_{S_k}.
\end{equation}

Here, the indices $\{i,j\}$ pertain to the redshift bins for the distance modulus, $S_k$ represents the magnitude of the systematic error, and $\sigma_{S_k}$ represents its standard deviation uncertainty.

\subsection{OHD+Pantheon+CMB}

In our study, we have additionally analyzed the CMB shift parameter data, obtained from observations by the Planck 2018 results \cite{Planck/2020}, to evaluate its influence on the current constraints of our model. In this scenario, the shift parameters $\mathcal{R}$ and $\ell_a$ are calculated as,
\begin{gather}
\mathcal{R}=\sqrt{\Omega_{m0}H_0^2}r(z_\star)/c,\\
\ell_a=\pi r(z_\star)/r_s(z_\star),
\end{gather}
where $r_{s}(z)$ represents the comoving sound horizon at redshift $z$, and $z_\star$ is the redshift to the photon-decoupling surface, which can be calculated using a fitting formula \cite{Hu/1996,Zhai}. We rely on the estimates derived from the Planck 2018 data, as detailed in \cite{Zhai}, using the data vector and covariance data specified in Eq. (31) of the referenced work.

Now, we derive constraints on the parameters of our cosmological model with a time-varying $\Lambda(t)$-term across all three cases using the combined OHD+Pantheon+CMB dataset. This was achieved by minimizing the total chi-squared function,
\begin{equation}
\chi^{2}_{total} = \chi^{2}_{OHD} + \chi^{2}_{Pantheon}+\chi^{2}_{CMB}.
\end{equation}

\begin{figure}[h]
\centering
\includegraphics[scale=0.6]{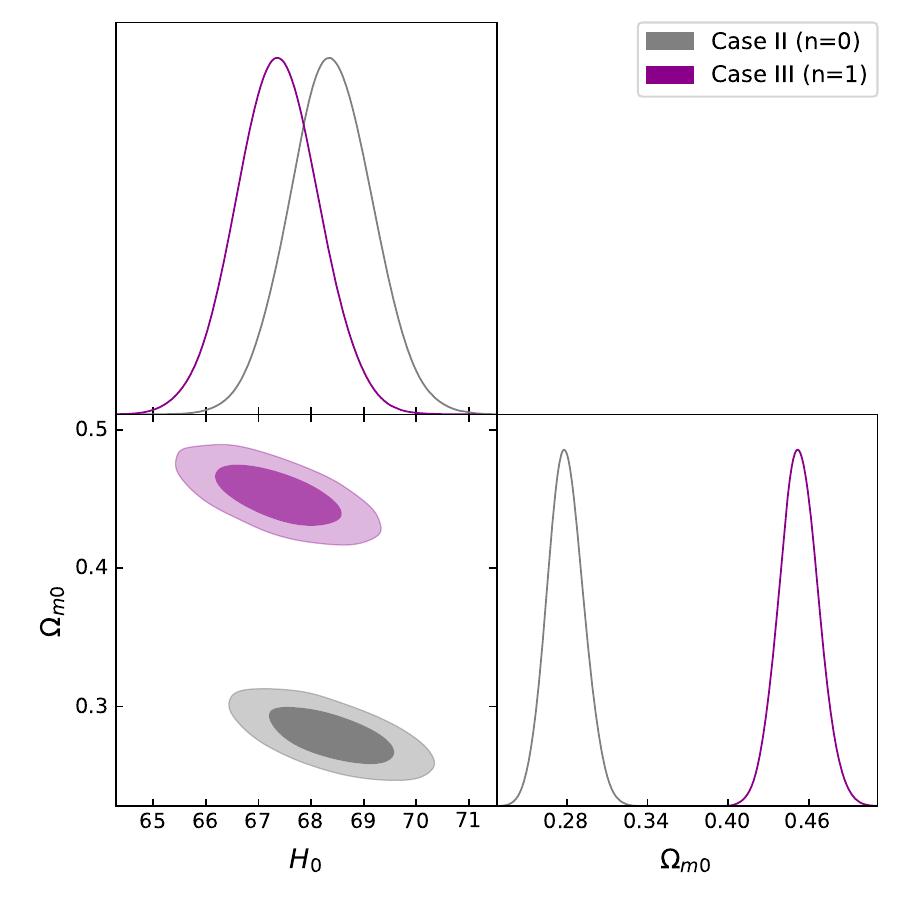}
\caption{The constraints on the model parameters for Cases II and III are presented at the $1-\sigma$ and $2-\sigma$ confidence intervals, derived from the combined OHD+Pantheon+CMB dataset.}
\label{F_01}
\end{figure}

The constraints obtained for the model parameters apply to all three cases for the combined OHD+Pantheon+CMB dataset, as presented in Tab. \ref{tab}. Furthermore, the $1-\sigma$ and $2-\sigma$ likelihood contours for the model parameters are depicted in Figs. \ref{F_n} and \ref{F_01}.

\subsection{Analysis of model selection and information criteria}

In this subsection, we will examine different statistical criteria and methods for selecting models. To do this, we will employ the Akaike information criterion (AIC) \cite{AIC1} and the Bayesian information criterion (BIC) \cite{AIC2} to evaluate various models' predictive performance using datasets. The AIC addresses the problem of model adequacy by serving as an estimator of Kullback-Leibler information, exhibiting asymptotically unbiased properties. The AIC estimator, under the assumption of Gaussian errors, is defined as \cite{AIC3, AIC4},
\begin{equation}
AIC = -2 \ln{(\mathcal{L}_{max})}+2 k + \frac{2 k\, (k+1)}{N_{tot}-k-1}.
\end{equation}

In this context, $k$ signifies the number of parameters in the model, $\mathcal{L}_{max}$ represents the maximum likelihood value of the datasets under examination, and $N_{tot}$ denotes the total number of data points. For extensive datasets, the formula simplifies to $AIC\equiv -2 \ln{(\mathcal{L}_{max})}+2 k $ \cite{AIC5}, a modified AIC criterion that is universally applicable. In addition, the BIC, functioning as a Bayesian evidence estimator \cite{AIC4,AIC5,AIC6}, is formulated as
\begin{equation}
BIC= -2 \ln{(\mathcal{L}_{max})}+ k \log(N_{tot}).  
\end{equation}

Here, $\chi^2_{min}=-2 \ln{(\mathcal{L}_{max})}$. In our effort to rank a series of competing models according to their fitting quality to observational data, we utilize the previously mentioned criteria. Our focus is particularly on the relative difference in Information Criterion (IC) values within the set of models under consideration. This difference, denoted as $\Delta IC_{model}= IC_{model}-IC_{min}$, compares the IC value of each model to the minimum IC value among the competing models. Finally, according to Jeffreys scale \cite{AIC7}, if $\Delta IC \leq 2$, the model is statistically consistent with the most favored model by the data. A range of $2 < \Delta IC < 6$ implies a moderate tension between the two models, while $\Delta IC \geq 10$ indicates a significant tension.

In our analysis of these tests, we have focused on Case II ($n=0$), which corresponds to the $\Lambda$CDM model as a baseline. We then compared this case with Cases I and III. The results of the model selection analysis are shown in Tab. \ref{tab}. These results indicate that for Case I, the differences are $\Delta AIC= 0.78$ and $\Delta BIC= 0.25$, whereas for Case III, the differences are $\Delta AIC= 0.03$ and $\Delta BIC= 0.03$. Therefore, the value of $\Delta IC$ is less than 2 for each comparison, indicating that both Case I and Case III are compatible not only with the $\Lambda$CDM model (Case II) but also with the observational datasets.

\begin{widetext}

\begin{table*}[h]
\begin{center}
%\adjustbox{width=0.5\textwidth}{
\begin{tabular}{l c c c c c c c c}
\hline\hline 
Cases              & $H_{0}$ $(km/s/Mpc)$ & $\Omega_{m0}$ & $n$ & $q_{0}$ & $z_{tr}$ & $\chi^2_{min}$ & $AIC$ & $BIC$ \\
\hline
Case I ($n$ is free) & $67.8^{+1.7}_{-1.7}$  & $0.37^{+0.12}_{-0.12}$  & $0.56^{+0.64}_{-0.73}$ & $-0.44^{+0.14}_{-0.09}$ & $0.79^{+0.17}_{-0.09}$ &$1439.75$&$1445.75$&$1448.85$\\

Case II ($n=0$) & $68.4^{+1.6}_{-1.6}$ & $0.279^{+0.027}_{-0.026}$  & $-$ & $-0.58^{+0.03}_{-0.02}$ & $0.74^{+0.25}_{-0.24}$&$1442.53$&$1446.53$&$1448.6$ \\

Case III ($n=1$) & $67.4^{+1.6}_{-1.6}$ & $0.453^{+0.029}_{-0.029}$  & $-$ & $-0.32^{+0.09}_{-0.09}$ & $0.82^{+0.37}_{-0.37}$&$1442.5$&$1446.5$&$1448.57$ \\

\hline\hline
\end{tabular}
%}
\caption{The table presents the constraints on the model parameters for all three cases based on the combined OHD+Pantheon+CMB dataset and information criteria.}
\label{tab}
\end{center}
\end{table*}

\begin{figure}[h]
\centering
\includegraphics[scale=0.85]{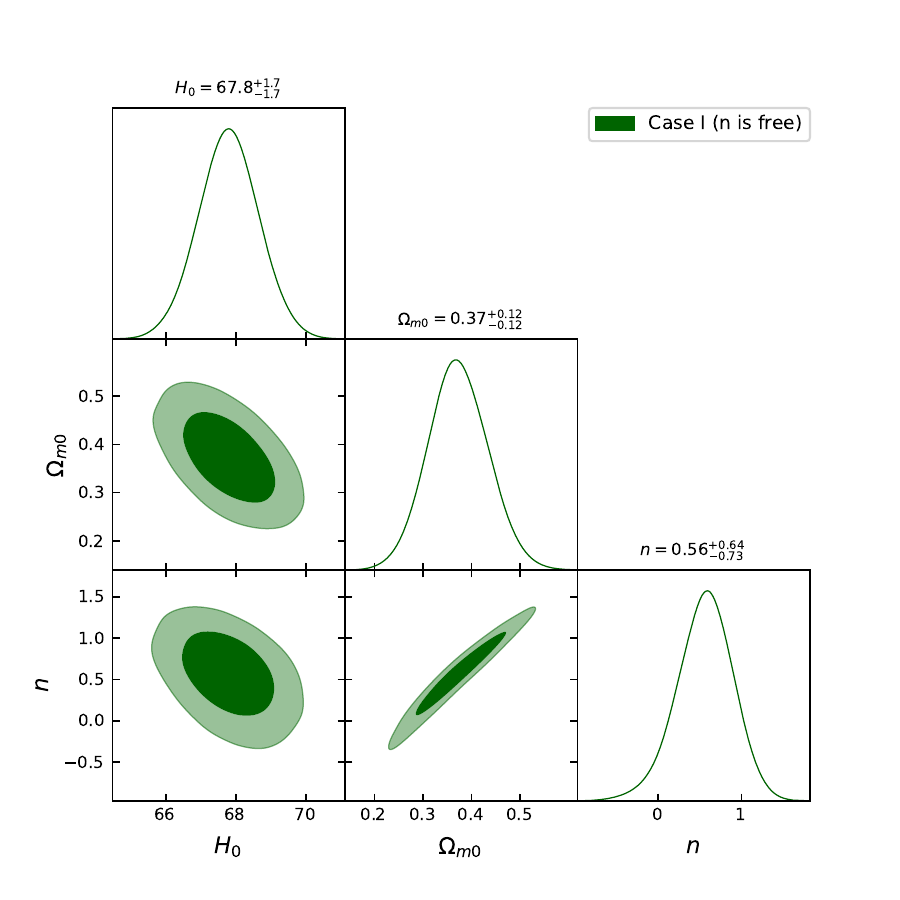}
\caption{The constraints on the model parameters for Case I are presented at the $1-\sigma$ and $2-\sigma$ confidence intervals, derived from the combined OHD+Pantheon+CMB dataset.}
\label{F_n}
\end{figure}
    
\end{widetext}

\section{Cosmological Parameters} \label{sec5}

In this section, we delve into the physical characteristics of established cosmological parameters, including the deceleration parameter, energy density, total EoS parameter, jerk parameter, and $Om(z)$ diagnostic.

\subsection{The deceleration parameter}
Based on cosmological observational data, cosmic acceleration is a relatively recent development. To comprehensively grasp the universe's evolutionary trajectory, a cosmological model should account for both the decelerating and accelerating phases of expansion \cite{Mamon1,Mukherjee1}. Consequently, it is vital to analyze the behavior of the deceleration parameter $q$ to gain a holistic understanding of cosmic evolution. The deceleration parameter, expressed as a function of the Hubble parameter $H$, is defined as
\begin{equation} \label{q}
q=-1-\frac{\dot{H}}{H^2}.
\end{equation}

From Eqs. \eqref{Hz} and \eqref{q}, we obtain
\begin{equation}
    q(z)=-1-\frac{\gamma \Omega_{m0} (1+z)^{\gamma }}{(n-2) \left(\Omega_{m0} \left((1+z)^{\gamma }-1\right)+1\right)}.
\end{equation}

In Fig. \ref{F_q}, the behavior of the deceleration parameter $q(z)$ exhibits a signature flip for all three cases: when $n$ is a free parameter, when $n=0$, and when $n=1$. The transition redshift $z_{tr}$ denotes the epoch when the universe's expansion transitioned from a matter-dominated decelerating ($q>0$) to accelerating ($q<0$) in the recent past. In addition, we anticipate that the universe will ultimately evolve towards a de Sitter expansion ($q=-1$) at lower asymptotic redshifts. Our analysis indicates that the present value of the deceleration parameter ($q_0$) and transition redshift ($z_{tr}$) (as presented in Tab. \ref{tab}) aligns well with observational data \cite{Hernandez,Basilakos,Roman,Jesus,Cunha}. This agreement reinforces the compatibility of our cosmological model with empirical observations.

\subsection{The total EoS parameter}

The total EoS parameter describes the relationship between total energy density and total isotropic pressure in the universe. It plays a crucial role in defining the different phases of cosmic evolution. For instance, in the EoS parameter, the dust phase is characterized by $\omega_{tot}=0$, while $\omega_{tot}=\frac{1}{3}$ corresponds to the radiation-dominated phase. The vacuum energy, represented by $\omega_{tot}=-1$ aligns with the $\Lambda$CDM model and signifies a phase of accelerated expansion. Recent discussions in cosmology have focused on the accelerating phase of the universe, which is associated with $\omega_{tot}<-\frac{1}{3}$. This range includes the quintessence phase ($-1<\omega_{tot}<-\frac{1}{3}$) and the phantom regime ($\omega_{tot}<-1$). These phases represent scenarios where the universe's expansion accelerates at an increasing rate, with the phantom regime suggesting an even more rapid acceleration. 

\begin{figure}[h]
   \begin{minipage}{0.48\textwidth}
     \centering
     \includegraphics[width=8.9cm,height=5.5cm]{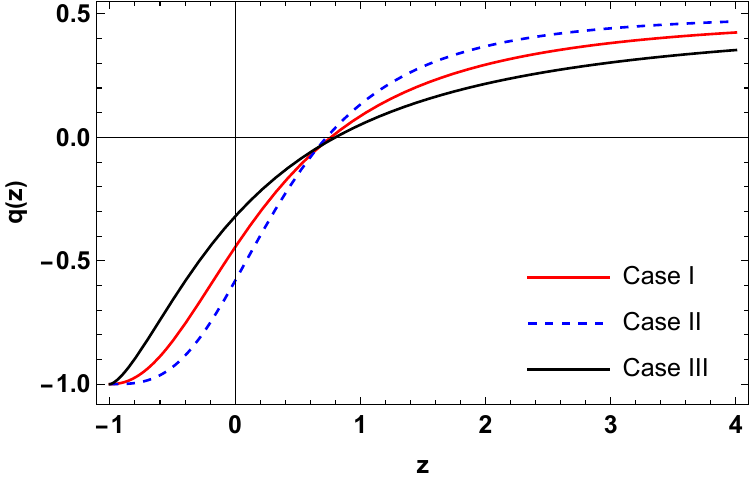}
     \caption{The plot illustrates the deceleration parameter corresponding to the observational constraints obtained from the combined OHD+Pantheon+CMB dataset.}\label{F_q}
   \end{minipage}\hfill
   \begin{minipage}{0.48\textwidth}
     \centering
     \includegraphics[width=8.9cm,height=5.5cm]{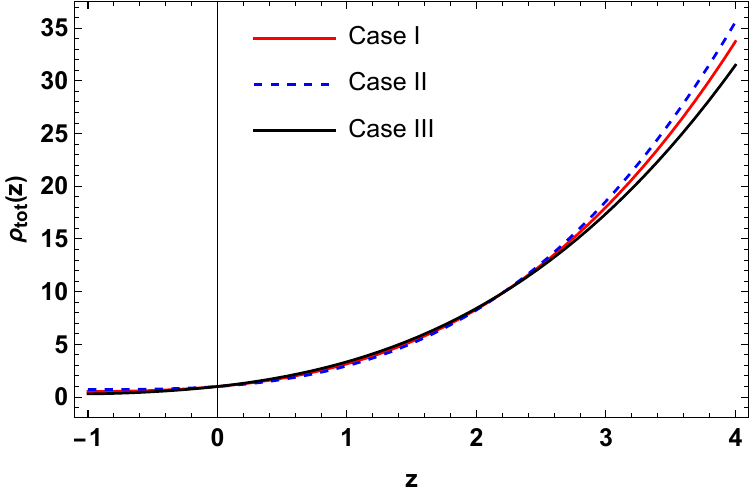}
     \caption{The plot illustrates the total energy density ($\rho_{tot}=\rho_{m}+\rho_{\Lambda}$) corresponding to the observational constraints obtained from the combined OHD+Pantheon+CMB dataset.}\label{F_rho}
   \end{minipage}
\end{figure}

\begin{figure}[h]
   \begin{minipage}{0.48\textwidth}
     \centering
     \includegraphics[width=8.9cm,height=5.5cm]{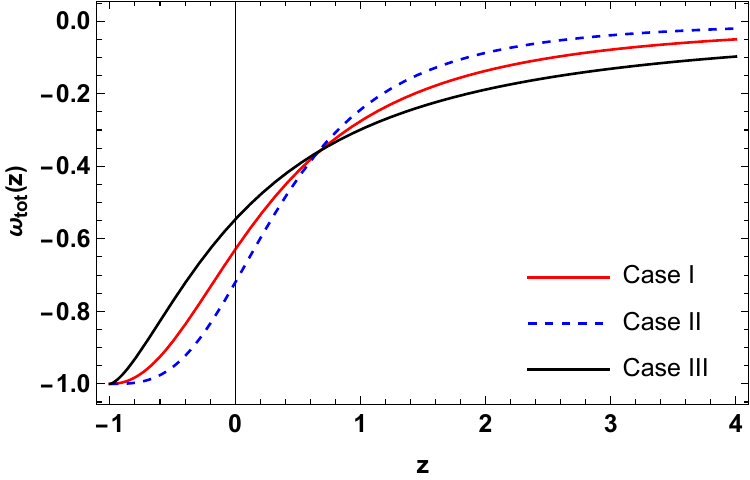}
     \caption{The plot illustrates the total EoS parameter corresponding to the observational constraints obtained from the combined OHD+Pantheon+CMB dataset.}\label{F_EoS}
   \end{minipage}\hfill
   \begin{minipage}{0.48\textwidth}
     \centering
     \includegraphics[width=8.9cm,height=5.5cm]{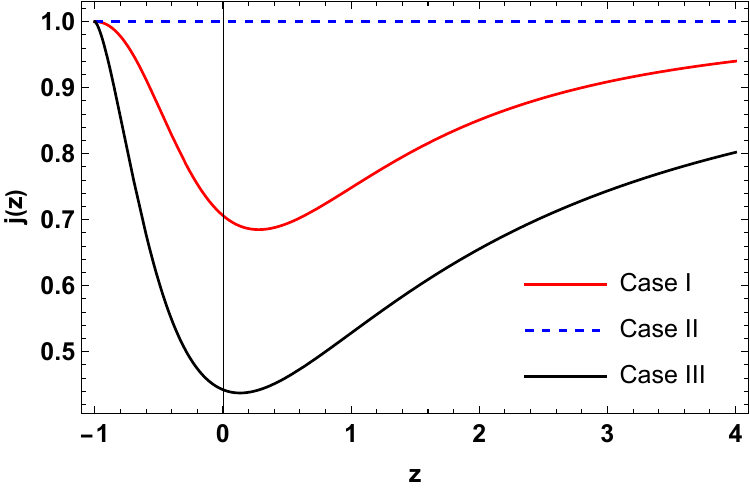}
     \caption{The plot illustrates the jerk parameter corresponding to the observational constraints obtained from the combined OHD+Pantheon+CMB dataset.}\label{F_j}
   \end{minipage}
\end{figure}

By employing Eqs. (\ref{EoS_tot}) and (\ref{Hz}), we derive the expression for the total EoS parameter $\omega_{\text{tot}}$. Fig. \ref{F_rho} depicts the total energy density's behavior, demonstrating a gradual decrease over time with a positive trend. This trend suggests that energy density will eventually approach zero in the distant future. In addition, the total EoS parameter, as depicted in Fig. \ref{F_EoS}, exhibits a similar evolution to the deceleration parameter, trending towards negative values at lower redshifts across all three cases for $n$. At present, the total EoS parameter lies within the quintessence regime ($-1<\omega_{tot}<-\frac{1}{3}$), indicating a phase of accelerated expansion. As redshift decreases, it tends towards the cosmological constant ($\omega_{tot}=-1$), suggesting a transition to a phase of exponential expansion. The present values of $\omega_{tot}$ are $-0.63^{+0.07}_{-0.06}$, $-0.72^{+0.02}_{-0.01}$, $-0.55^{+0.08}_{-0.08}$ for the cases where $n$ is a free parameter, $n=0$, and $n=1$, respectively \cite{Gruber}. These values align with the quintessence regime, indicating an evolving universe with a transitioning EoS parameter.

\subsection{The jerk parameter}

The jerk parameter is a cosmological quantity that extends the description of the universe's expansion rate beyond the conventional parameters of $H(t)$ and $q(t)$. It is derived from the fourth term in a Taylor series expansion of the scale factor a around a specific time $t_0$ \cite{Sahni:2002fz},
\begin{equation}
\begin{aligned}
\frac{a(t)}{a_0} & = 1+H_0(t-t_0)-\frac{1}{2}q_0H^2_{0}(t-t_0)^2+
\frac{1}{6}j_0H^3_{0}(t-t_0)^3 \\ & +O\left[(t-t_0)^4\right].
\end{aligned}
\end{equation}

The jerk parameter, represented by $j$, is defined as the third derivative of the scale factor with respect to cosmic time \cite{Visser:2004bf},
\begin{equation}
j=\frac{1}{a}\frac{d^3a}{d\tau
^3}\left[\frac{1}{a}\frac{da}{d\tau}\right]^{-3}=q(2q+1)+(1+z)\frac{dq}{dz}.
\end{equation}

The jerk parameter plays a crucial role in understanding cosmic dynamics, especially in the context of various DE models \cite{Visser:2004bf}. Its value can help establish a connection between these models and standard universe models, aiding in the search for an appropriate candidate to describe cosmic dynamics. For instance, in the flat $\Lambda$CDM model (Case II), the jerk parameter is assigned a value of $j=1$. Fig. \ref{F_j} illustrates the evolution of the jerk parameter for all three cases of $n$. The results indicate that in cases I ($n$ is free) and III ($n=1$), there is a deviation from the flat $\Lambda$CDM model (Case II with $n=0$) based on the best-fit values. Observations suggest that in these models, the current value of $j$ exceeds $1$ for Cases I and III ($j=0.71$ and $j=0.44$, respectively) \cite{Mamon2}. Therefore, in the current models where $n$ is a free parameter and $n=1$ (where $j_0>0$ and $q_0<0$), it is evident that the dynamic DE model under consideration is the most likely explanation for the current acceleration. This outcome suggests the need for further investigation into the dynamic DE model to better understand its implications for cosmic acceleration.

\subsection{$Om(z)$ diagnostics}

The $Om(z)$ diagnostic is an additional valuable tool for categorizing different DE cosmological models \cite{Sahni}. Its simplicity lies in its reliance solely on the first-order derivative of the scale factor, making it particularly straightforward to apply and interpret in cosmological studies. For a spatially flat universe, its expression is given by
\begin{equation}
Om\left( z\right) =\frac{\left( \frac{H\left( z\right) }{H_{0}}\right) ^{2}-1%
}{\left( 1+z\right) ^{3}-1}.    
\end{equation}

\begin{figure}[h]
\centering
\includegraphics[scale=0.7]{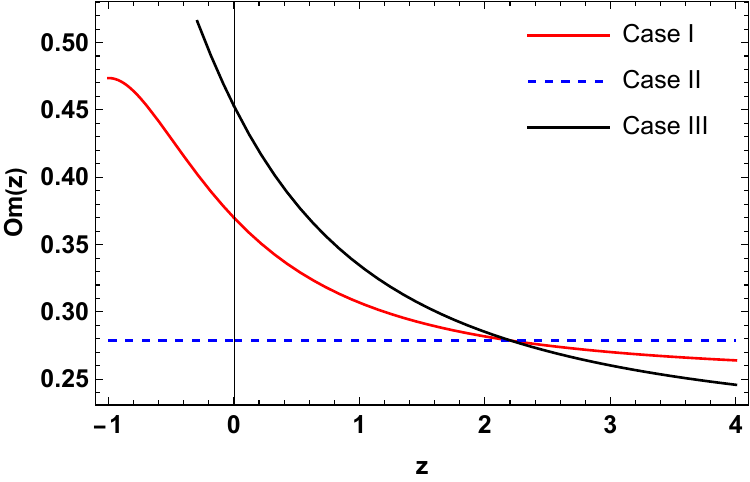}
\caption{The plot illustrates the $Om(z)$ diagnostic corresponding to the observational constraints obtained from the combined OHD+Pantheon+CMB dataset.}
\label{F_Om}
\end{figure}

In this context, $H_0$ denotes the present value of the Hubble parameter. A negative slope of $Om(z)$ signifies quintessence-type behavior, indicating a slowing rate of expansion over time, while a positive slope corresponds to phantom behavior, signifying an accelerating rate of expansion over time. A constant $Om(z)$ value indicates the $\Lambda$CDM model. Fig. \ref{F_Om} demonstrates that the $Om(z)$ diagnostic, across its evolution for the three cases of $n$ shows a negative slope for Cases I and III, and a constant value for Case II. Therefore, according to the $Om(z)$ test, we can infer that our $\Lambda(t)$CDM model exhibits quintessence-type behavior.

\section{Conclusion} \label{sec6}

Cosmology has grappled with two significant challenges: dark matter and DE. Dark matter, which does not interact electromagnetically but exerts gravitational effects, remains undetected as a particle in the standard model of particle physics \cite{CDMS,Akerib}. The phenomenon of DE, indicated by the universe's observed accelerated expansion, presents another puzzle. While the cosmological constant in GR can account for this acceleration, the fine-tuning and coincidence problems associated with the cosmological constant have motivated the exploration of alternative explanations for DE. Consequently, numerous DE models have emerged in an attempt to address these issues and provide a more comprehensive understanding of the universe's dynamics. In addition, theories involving time-varying $\Lambda(t)$ or vacuum decay have been proposed to explain the effects attributed to dark matter, adding further complexity to the quest for a complete cosmological model \cite{IE}. 

In this manuscript, we have investigated signature flips within cosmological models featuring a time-varying vacuum energy term $\Lambda(t)$, specifically using the power-law form $\Lambda = \alpha H^n$. Through the application of the Markov Chain Monte Carlo technique to analyze OHD (31 points), Pantheon data (1048 points), and CMB data, we have constrained the model parameters under three scenarios (see Tab. \ref{tab} and Figs. \ref{F_n} and \ref{F_01}: when $n$ is a free parameter (Case I), when $n = 0$ (Case II), and when $n = 1$ (Case III). Then, we have analyzed the behavior of various cosmological parameters across all three cases, encompassing the deceleration parameter, energy density, total EoS parameter, jerk parameter, and the $Om(z)$ diagnostic.

From Fig. \ref{F_q}, we observed a signature flip in the behavior of the deceleration parameter across all three cases (when $n$ is free, $n=0$, and $n=1$). This transition occurred at the redshift $z_{tr}$, indicating a shift from a matter-dominated decelerating phase to an accelerating phase in the recent past. Our model's present values of $q_0$ and $z_{tr}$ align well with observational data, supporting the model's consistency with empirical observations. In addition, the total energy density exhibited a gradual decrease over time, with a trend towards zero in the distant future (see Fig. \ref{F_rho}). The total EoS parameter in Fig. \ref{F_EoS} showed a similar evolution to the deceleration parameter, indicating an accelerated expansion phase at present and a transition towards a cosmological constant in the future. The jerk parameter's behavior deviated from the flat $\Lambda$CDM model (Case II) in Cases I and III (see Fig. \ref{F_j}), indicating a dynamic DE model as the most likely explanation for the current acceleration. This deviation suggests the need for further investigation into the dynamic DE model's implications for cosmic acceleration. Further, the $Om(z)$ diagnostic in Fig. \ref{F_Om} exhibited a negative slope for Cases I and III, indicating quintessence-type behavior for our $\Lambda(t)$CDM model.

Finally, our analysis suggests that the time-varying vacuum energy term $\Lambda(t)$ provides a viable framework for describing the observed dynamics of the universe. The model's compatibility with observational data and its ability to explain key phenomena such as cosmic acceleration highlight its potential as a valuable tool for understanding the evolution of the cosmos. Further studies and refinements of this model could lead to deeper insights into the fundamental nature of dark energy and its role in shaping the universe's history.

\section*{Acknowledgment}
This research was funded by the Science Committee of the Ministry of Science and Higher Education of the Republic of Kazakhstan (Grant No. AP22682760).

\section*{Data Availability Statement}
This article does not introduce any new data.

%%%%%%%%%%%%%%%%%%%%%%%%%%%%%%%%%%%%%%%%%%%%%%%%%%%%%%%%%%%%%%%%%%%%%%%%%%%%%%%%%
%%
%%

\end{document}